\documentclass[aps, prl, twocolumn,showpacs,floatfix]{revtex4}
\usepackage{graphicx,epsfig,amsmath}

\newcommand{\subfig}[2]{Fig.~\ref{fig:#1}(#2)} 

\newcommand{\Vds}{\mbox{$\text{V}_{\text{ds}}$}}
\newcommand{\Vexc}{\mbox{$\text{V}_{\text{exc}}$}}
\newcommand{\Vg}{\mbox{$\text{V}_{\text{g}}$}}

\newcommand{\Vb}{\mbox{$\text{V}_{\text{b}}$}}
\newcommand{\Vc}{\mbox{$\text{V}_{\text{c}}$}}
\newcommand{\Va}{\mbox{$\text{V}_{\text{a}}$}}
\newcommand{\Vt}{\mbox{$\text{V}_{\text{t}}$}}

\newcommand{\eSqOnh}{\mbox{$\text{e}^{\text{2}}/\text{h}$}}

\begin{document}

\title{Few-electron quantum dots in III-V ternary alloys: role of fluctuations}


\author{G.~Granger}
	\email{Ghislain.Granger@nrc.ca}	
\author{S.A.~Studenikin}
\author{A.~Kam}
\author{A.S.~Sachrajda}
\author{P.J.~Poole}

\affiliation{Institute for Microstructural Sciences, National Research Council Canada, Ottawa, ON, Canada, K1A 0R6}


\begin{abstract}

We study the electron transport properties of gated quantum dots formed in InGaAs/InP and InAsP/InP quantum well structures grown by chemical-beam epitaxy. For the InGaAs quantum well, quantum dots form directly underneath narrow gates due to potential fluctuations. We measure the Coulomb-blockade diamonds in the few-electron regime of a single dot and observe photon-assisted tunneling peaks under microwave irradiation. A singlet-triplet transition at high magnetic field and Coulomb-blockade in the quantum Hall regime are also observed. For the InAsP quantum well, an incidental triple quantum dot forms due to potential fluctuations. Tunable quadruple points are observed via transport measurements.

\end{abstract}


\maketitle



Materials with large spin-orbit coupling, such as InGaAs, are currently of considerable interest due to potential spintronics applications. Although lateral quantum dots (QDs) are routinely demonstrated in GaAs, the InGaAs material has not been suitable for this technology because of poor quality Schottky gates. Very recently, two groups have applied the split-gate technique on InGaAs structures using dielectrics between the gates and the heterostructure \cite{Sun2010, Deon2010}. Pre-patterned InP substrate technology allows us to substantially reduce leakage and exclude dielectrics, i.e.~use Schottky gates rather than MOS gates. In addition, this technology is promising for optical applications and g-factor engineering \cite{Granger2009}.

Here we continue developing this technology and study the formation of single QDs in the few-electron regime in an InGaAs quantum well [QW] under a narrow gate. We find that potential fluctuations play an important role in the formation of the QDs, as it was the case for incidental dots in GaAs \cite{Nicholls1993, Weis1992, Gaudreau2006}. We perform a comprehensive study of these dots using standard techniques. A case of transport measurements through an incidental triple quantum dot formed in an InAsP QW is also demonstrated.


\begin{figure}[hbt]
\setlength{\unitlength}{1cm}
\begin{center}
\begin{picture}(8,7.6)(0,0)
\includegraphics[width=8cm, keepaspectratio=true]{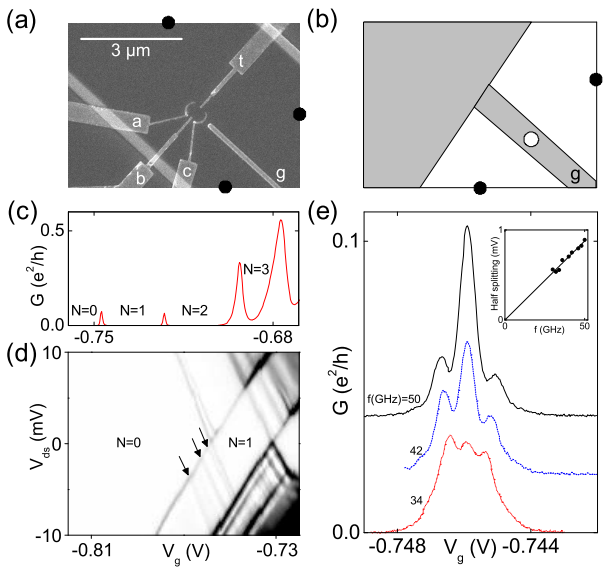}
\end{picture}
\end{center}
\caption{(Color online) (a) Electron micrograph of a split gate device nominally identical to the ones measured. Black dots indicate Ohmic contacts here and in Figs.~\ref{fig:1}(b) and \ref{fig:3}(a). TiAu gate electrodes, labeled a, b, c, t, and g, climb on the top facet of the 5~$\mu$m-wide active region (the side walls of the ridge structure appear light grey). (b) Close up schematic showing how a quantum dot can form inside a fluctuation underneath gate g. The grey area in the upper left corner is depleted of electrons, as the voltages on the other gates are passed pinch-off. (c) Coulomb-blockade in the few-electron regime. The excitation voltage \Vexc=20~$\mu$V. (d) dI/d\Vds~in the \Vds-\Vg~plane showing Coulomb-blockade diamonds. White (black) is low (high) dI/d\Vds. Arrows point to lines from mesoscopic modulations of the density of states in the leads. (e) Dependence of the first Coulomb charging peak on microwave frequency. Inset: half the splitting between the satellite peaks vs.~microwave frequency and linear fit.}
\label{fig:1}
\end{figure}

We start with a sample containing an InGaAs QW in InP, 70~nm below the surface. NiAuGe Ohmic contacts are located on the edges of a 5~$\mu$m-wide active region. The split gate layout, similar to that of Ref.~\cite{Ciorga2000}, is shown in \subfig{1}{a}. Several devices had similar QD characteristics, and here we present one of them in which we have achieved the one-electron regime. As shown schematically in \subfig{1}{b}, the QD forms directly under 160~nm-wide gate g when the gate voltage \Vg~is brought close to the pinch-off voltage. The remaining gates are biased passed their pinch-off to deplete the region in the upper left corner of \subfig{1}{b}. As the electron density under gate g gradually decreases to zero, a small puddle of electrons survives as a result of a fluctuation in the local QW thickness or composition in a similar way as reported in \cite{Nicholls1993, Weis1992}. We measure conductance G as a function of \Vg~at 0.3~K using standard lock-in techniques. After illumination, the electron density in the ungated areas is 6$\times10^{15}$~m$^{-2}$, and the mobility is $\gtrsim$2~m$^2$/Vs, as estimated from two-terminal measurements.


Figure~\ref{fig:1}(c) shows that Coulomb-blockade peaks occur in the conductance as \Vg~is swept. The peak positions change very little when the voltage on any other gate is changed (ratio $\sim$1/30, not shown), which confirms that the QD is formed under gate g. The Coulomb-blockade peaks in \subfig{1}{c} correspond to charging of the last potential fluctuation.
 
In order to confirm that we operate in the few-electron regime, we apply a drain-source bias, \Vds, and observe typical Coulomb-blockade diamonds in dI/d\Vds~(\subfig{1}{d}). We have determined from measurements where $|$\Vds$|$ goes up to 35~mV and \Vg~down to -0.85~V (not shown) that the last diamond never closes. We conclude that we have completely emptied the dot. We also observe faint lines that terminate in the N=0 region (arrows in \subfig{1}{d}). These are not excited states of the QD. Detailed studies confirm that they originate from mesoscopic modulations in the density of states in the leads \cite{Lim2009, Escott2010} and completely disappear if a 1~T perpendicular magnetic field is applied. The size of the first Coulomb-blockade diamond gives a charging energy of 6~meV, which corresponds to a dot diameter of $\approx$60~nm. The slopes of the diamond edges allow the extraction of the dimensionless constant $\alpha_g$=0.247, which converts the \Vg~axis into electrochemical potential. The spacings between the first few Coulomb-blockade peaks in \subfig{1}{c} are consistent with a shell structure filling. Indeed, the N=2 valley (corresponding to a filled \textit{s} shell of the artificial atom) is larger than both the N=1 and N=3 valleys.

To confirm $\alpha_g$ from an independent experiment, we shine microwaves on the QD. We measure the evolution of the first Coulomb charging peak from \subfig{1}{c}, which survives even at 6~K (not shown), in the presence of a cw microwave irradiation. The results are in \subfig{1}{e}. At low power ($\sim\mu$W), we find a pair of satellites on the sides of the conductance peak, as shown in \subfig{1}{e}. At higher powers, multiple-photon absorption is observed, so an additional pair of satellites develop further away from the Coulomb-blockade peak (not shown). In order to confirm that the satellites originate from PAT, we repeat the experiment at several frequencies f between 30 and 50~GHz. We fit each trace to a sum of three Lorentzians (not shown) and extract the spacing between the satellites. The splitting is linear in f over the entire frequency range (\subfig{1}{e} inset), which provides an independent measurement of $\alpha_g$=0.236 that compares well with that extracted from Coulomb-blockade diamonds. The fact that two satellites are observed signifies that the QD is symmetrically coupled to both leads. An example of PAT with QDs in GaAs can be found in Ref.~\cite{Oosterkamp1997}.

\begin{figure}[hbt]
\setlength{\unitlength}{1cm}
\begin{center}
\begin{picture}(8,5.9)(0,0)
\includegraphics[width=8cm, keepaspectratio=true]{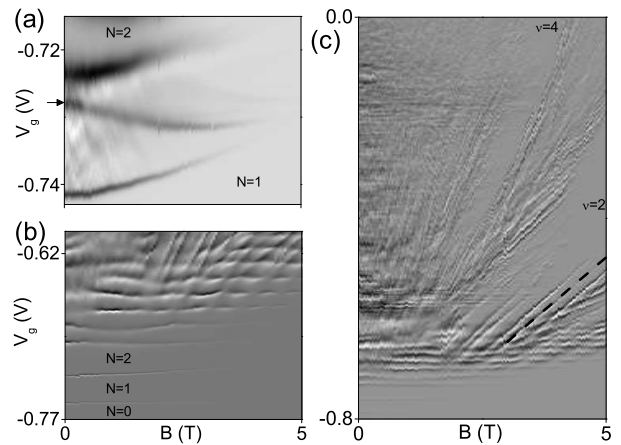}
\end{picture}
\end{center}
\caption{(a) dI/d\Vds~at \Vds=4mV vs. B where the line due to the triplet excited state is indicated by an arrow. White (black) is low (high) dI/d\Vds. (b) B-dependence of dG/d\Vg~in the few-electron regime. Black (white) is negative (positive) dG/d\Vg.  (c) dG/d\Vg~map showing charging effects in the QHE regime. The values of $\nu$ indicated are the average filling factors in the 2DEG directly under the gate. White (black) is negative (positive) dG/d\Vg. The dashed line is an example of charging feature near $\nu$=2.}
\label{fig:2}
\end{figure}

We now examine how the conductance evolves when a magnetic field, B, perpendicular to the two-dimensional electron gas (2DEG) is applied. Figure~\ref{fig:2}(a) shows the evolution of the Coulomb charging peaks between N=1 and N=2 from \subfig{1}{c} as a function of B at \Vds=4~mV. The presence of a nonzero bias across the dot results in a current stripe between the N=1 and N=2 regions. The extra line indicated by an arrow in \subfig{2}{a} within the current stripe originates from transport through an excited state. For a dot with two electrons, the exchange energy between them results in a singlet ground state and a triplet excited state, hence the excited state feature seen in \subfig{2}{a} is ascribed to the triplet. At B=0~T, the singlet-triplet gap is $\epsilon_{\text{ST}}\approx$3~ meV (i.e. $\sim$5 times larger than in GaAs dots \cite{Kyriakidis2002}). Transport through the triplet state occurs at smaller \Vg~as B grows, and a singlet-triplet transition occurs at B$\sim$5~T. However, the signal decreases significantly at such high fields, so it is difficult to resolve the transition itself and to determine whether there is an avoided crossing. We note that a singlet-triplet transition in a few-electron lateral InGaAs/InAlAs QD has been reported in Ref.~\cite{Deon2010}.

To investigate the role of fluctuations at larger electron numbers, we look at a large overview of dG/d\Vg~in the \Vg-B plane. The results are shown \subfig{2}{b} and (c). In \subfig{2}{b}, the Coulomb-blockade peaks curve up as a function of B. As the electron number increases, the conductance peak amplitude may contain information about the change between the singlet and spin-polarized phases \cite{Ciorga2002}. Amplitude modulations are seen in \subfig{2}{b}; however, charging effects that will be described below play an important role in our case.

Figure~\ref{fig:2}(c) shows the dG/d\Vg~diagram all the way up to \Vg=0. Such a large range of \Vg~allows one to study what happens as the average electron density under the gate increases. Flat regions, corresponding to plateaus in G, are seen whenever the electron density under the gate is such that the filling factor $\nu$ is constant. The cases of $\nu$=4 and 2 are indicated in \subfig{2}{c}. In addition to flat regions, there are sets of parallel lines in \subfig{2}{c} that correspond to charging of isolated compressible regions arising from potential fluctuations in the Quantum Hall Effect (QHE) regime. Similar charging effects have also been observed in Si MOSFETs \cite{Cobden1999} and graphene flakes \cite{Branchaud2010, Velasco2010} using transport, and in GaAs and graphene using a scanning single-electron transistor \cite{Ilani2004, Martin2009}. The average vertical offset $\Delta$\Vg~between 5 lines (the dashed line in \subfig{2}{c} is an example) in the \Vg-B plane on the hole side of the $\nu$=2 region at B=4~T allows us to estimate the average charging energy U=$\alpha_g\Delta$\Vg=2~meV, where the value of $\alpha_g$ at B=4~T is 0.075 from a Coulomb-blockade diamond in the \Vds-\Vg~plane (not shown). Assuming a circular dot, the average fluctuation diameter in this regime is D=e$^2$/(4$\epsilon_{\text{InP}}$U)$\approx$190~nm, where $\epsilon_{\text{InP}}$=12$\epsilon_{0}$.

\begin{figure}[bt]
\setlength{\unitlength}{1cm}
\begin{center}
\begin{picture}(8,6.3)(0,0)
\includegraphics[width=8.0cm, keepaspectratio=true]{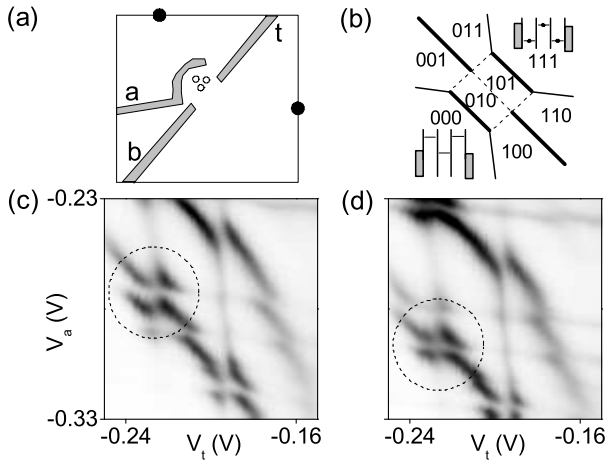}
\end{picture}
\end{center}
\caption{(a) Schematic showing how three split-gates are used together with potential fluctuations to form the triple quantum dot. The other gates from \subfig{1}{a} are grounded so they have no confining effect.  (b) Diagram explaining transport at QPs in the parallel TQD. The thick addition line becomes doubled at the QPs, because its location in voltage space depends on whether the other two dots are charged or not. The relative electronic configurations are indicated from an unknown (N$_1$,N$_2$,N$_3$). (c) and (d): Stability diagram of the conductance of the TQD showing several pairs of QPs (e.g. dashed circles). The sample was initially illuminated with a red LED so n=3.2$\times 10^{15}$~m$^{-2}$ and $\mu$$\geq$2~m$^2$/Vs. The greyscale is such that 0 (0.06~\eSqOnh) is white (black). \Vg=\Vc=0~V. \Vexc=50~$\mu$V.  (c) \Vb=-0.225~V (d) \Vb=-0.215~V.}
\label{fig:3}
\end{figure}

Now, we present results measured with an InAsP QW, where quantum dots are confined in the expected region defined by the split gates a, b, and t, as shown in the schematic in \subfig{3}{a} \cite{footnote1}. In this device, single-electron charging features respond equally to gates a, b, and t, which indicates that the QDs indeed are formed in the region confined by the three gates. The transport diagrams showing the conductance G as a function of \Va~and \Vt~for two different values of \Vb~are shown in \subfig{3}{c} and (d). It is seen that the charging lines have three different slopes, which is a signature of a triple quantum dot [TQD] \cite{Schroer2007, Gaudreau2009, Granger2010}. Three dots form because potential fluctuations break down the smooth potential built by the three split gates into three smaller dots. These QDs are coupled to both reservoirs, as transport in \subfig{3}{c} and (d) is not limited to quadruple points (QPs) as it is the case for a series TQD formed using 8 or 9 gates \cite{Schroer2007, Granger2010}. Nevertheless, transport is stronger where three lines intersect each other, as seen in \subfig{3}{c} and (d) for two values of \Vb~as indicated by the dashed circles. At the QPs, the addition line with intermediate slope is the most visible, which indicates that the corresponding dot is the most coupled to both leads. The fact that this line splits into two at the QPs before returning to a single line is explained schematically in \subfig{3}{b}; this occurs because of the capacitive coupling to the charges on the other two dots, consistent with the model calculations in Fig.~5 of Ref.~\cite{Granger2010}.



In summary, we have achieved a few-electron quantum dot in a sample with an InGaAs quantum well defined by a single gate. We have characterized the dot using several common transport phenomena, namely, Coulomb-blockade diamonds in the few-electron regime, transport through the N=2 triplet excited state, photon-assisted tunneling, and charging of potential fluctuations in the quantum Hall effect regime. In a sample with an InAsP quantum well, we have achieved quantum dots with the split-gate geometry, but fluctuations in the quantum well thickness or composition resulted in the formation of a triple quantum dot. The results in this paper confirm that fluctuations play an important role in quantum dot formation in gated III-V structures grown on pre-patterned InP substrates. Such fluctuations may be advantageous in forming smaller multiple dot devices with fewer gates.

\acknowledgments

We acknowledge L. Gaudreau, P. Hawrylak, and M. Korkusinski for discussions. G.G. acknowledges financial support from the CNRS-NRC collaboration, and A.S.S. acknowledges support from CIFAR.


\begin{thebibliography}{00}
\bibitem{footnote1} We also performed experiments with single QDs forming under a single gate as part of split-gate designs using InAsP QWs (not shown) with results similar to those with single dots forming under single gates in InGaAs QWs.
\end{thebibliography}
\end{document}